\documentclass[journal]{IEEEtran}

\usepackage[T1]{fontenc}
\usepackage[utf8]{inputenc}
\usepackage{amsmath, amssymb}
\usepackage{graphicx}
\usepackage{booktabs}
\usepackage{multirow}
\usepackage{cite}
\usepackage{algorithm}
\usepackage{algorithmic}
\usepackage{url}
\usepackage{xcolor}
\usepackage{tabularx}
\usepackage{array}
\usepackage{longtable} 
\usepackage{array}     
\usepackage{enumitem}  
\usepackage{verbatim}

\title{A Hodge-Based Framework for Service Operational Analysis in Serverless  Platforms}

\author{Gianluca Reali,~\IEEEmembership{Member,~IEEE,} and Mauro Femminella,~\IEEEmembership{Member,~IEEE}
\thanks{Authors are with the Department of Engineering, University of Perugia, 06125 Perugia, Italy, and with Consorzio Nazionale Interuniversitario per le Telecomunicazioni (CNIT), 43124 Parma, Italy.}
\thanks{This work was funded by the European Union under the Italian National Recovery and Resilience Plan (NRRP) of NextGenerationEU, partnership on “Telecommunications of the Future” (PE00000001 - program ``RESTART'') and National Innovation Ecosystem (ECS00000041 - program ``VITALITY'').} 
\thanks{Corresponding author: Gianluca Reali (gianluca.reali@unipg.it).}
}

\begin{document}

\maketitle

\begin{abstract}
In this paper we propose a method for analyzing services deployed in serverless platforms. These services typically consists of orchestrated functions that can exhibit complex and non-conservative information flows due to the interaction of independently deployed functions under coarse-grained control mechanisms.
We introduce a topological model of serverless services and make use of the Hodge decomposition to partition observed operational flows into locally correctable components and globally persistent harmonic modes.
Our analysis shows that harmonic flows naturally arise from different kind of interactions among functions and should be interpreted as structural properties of serverless systems rather than configuration errors.
We present a systematic methodology for analyzing inter-function flows and deriving actionable remediation strategies, including dumping effects to contain the effects of harmonic inefficiencies as an alternative to completely restructure the topological model of the service.
Experimental results confirm that the proposed approach can uncover latent architectural structures leading to inefficiencies.
\end{abstract}

\begin{IEEEkeywords}
Serverless, FaaS, Hodge Decomposition, Sagas.
\end{IEEEkeywords}

\section{Introduction}
This paper addresses the critical issues that typically emerge in the implementation and deployment of complex cloud services in serverless mode \cite{faast}. 
Despite the growing diffusion of increasingly refined orchestration technologies and the identification of a considerable number of design patterns, when the implementation of a large number of cloud services of multiple types is carried out in serverless mode \cite{10.1145/3508360}, unexpected problems that affect the correct provision of services in function-as-a-service (FaaS) platforms may occur \cite{yu2023following,saha2024survey,calavaro2025beyondcloud}. 

The occurrence of these problems derives from a series of concomitant factors, the simultaneous management of which could be complex. 
The first of these factors is the potentially large number of independent or loosely coupled components that are required to run applications. Essentially, what distinguishes serverless deployment from previous approaches is the structuring and implementation of application logic using a chain of elementary stateless functions. The execution of a service therefore consists in the execution of a concatenated series of elementary functions, possibly reusable for different applications, which live on the network only for the time necessary for the needs of the application. This way, the network footprint, the associated costs, the energy consumption, and the attack surface are minimized. These benefits, however, come at the cost of increased code complexity, difficult debugging, intricate orchestration of interacting functions, and additional latency \cite{calavaro2025beyondcloud}\cite{10572125} \cite{Mohanty}. 

This brings us to another critical factor. The complexities of the interactions between functions could generate unexpected loops that are beyond the control of the implementers. For example, if the execution of a function depends on the value of an environment variable, and at a point in the chain of functions this value is inadvertently altered, an uncontrolled circular process can be triggered. This behavior could generates unexpected delays and various malfunctions, the root cause of which is difficult to find. Another example is the use of so-called compensations in design patterns \cite{Heus} \cite{orkesCompensationTransaction}. They are actions that are performed to undo or counterbalance the effects of a previous action, when an atomic transaction cannot be used. For example, if an order is placed through an e-commerce application and the payment fails, it must be canceled through a chain of actions that include verifying account funds, confirming card details, contacting the bank for potential fraud restrictions and release of resources. Such compensations are necessary, functions are distributed, there is no global commit, and failures can happen. Therefore these compensations introduce reverse action loops and additional calls. From the point of view of the system, they generate the circulation of "work". 
When the application is built with a limited number of microservices, it can be assumed that these loops are adequately managed with design patterns, such as the \textit{circuit breaker}, which avoids the presence of uncontrolled loops. However, when applications require a large number of functions to be concurrent, these loops may occur uncontrollably, generating logical "holes" corresponding to continuous and circular flows of data. 

Another factor to consider is that serverless functions could be activated as needed, with activation latencies that are not present in traditional deployment. This is the so-called \textit{cold start} phenomenon. Strategies to avoid this phenomenon have been the subject of intense research recently \cite{10.1145/3366623.3368139,varghese,icccn25_femminella}. However, sometimes it is unavoidable, especially when running functions are horizontally scaled. When this happens, several problems can occur. For example, if a function A invokes B and the latter is in cold start, a timeout can occur. Then A tries again and B in the meantime starts, generating a double execution. This generates a loop (local curl in topological jargon), that needs to be detected and managed. Other issues related to the same phenomenon are related to consistency in the backend, cache consistency issues, and more.  

In this context, this paper shows how the use of algebraic techniques developed in the context of Topological Signal Processing (TSP) allow identifying operational criticalities in the execution of applications. Our approach starts from the Hodge theory, which states that on a compact Riemannian manifold each de Rham cohomology class admits a unique harmonic representative. Then we consider the discrete analogue for graphs, where combinatorial Hodge theory identifies simplicial cohomology with the space of discrete harmonic forms, defined via the graph Laplacian. In this way, we resort to the Hodge decomposition of edge flows on a graph into three orthogonal components, a gradient component, a curl component, and a harmonic component. \cite{Schaub2021Tutorial}. The last two are particularly interesting for the analysis of the execution of applications because they are directly associated with the local closure of the service sagas \cite{saga} and the presence of structural loops that generate inefficiencies, performance degradation and energy consumption problems. 
Therefore, the general contributions of the paper can be summarized as follows.
\begin{itemize}
    \item Identification of the possible categories of problems that afflict the execution of applications that implement cloud services, implemented in serverless mode.
    \item Proposal and analysis of a TSP-based model for the identification of the causes of the malfunctions observed in accessing serverless cloud services. 
\end{itemize} 
More specifically, through Hodge decomposition, we demonstrate that it is possible to perform (i) a diagnosis of the operation of the service, using the harmonic component to identify global structural problems of the system, (ii) do automatic debugging, identifying functions that participate in load cycles that cannot be eliminated locally, (iii)  designing an algorithm that highlights curl and harmonic components for different types of problems, and (iv) define a new metric based on the "harmonic stress" of the system. 
Hodge decomposition allows us to identify the "payload", routable, through the gradient component, the local loops, through the curl component, and the structural inefficient patterns, through the harmonic components \cite{Schaub2020,Schaub2021Tutorial}.  

The main conceptual result of the paper is that the operational and architectural causes of the inefficiencies may not be purely architectural. Two deployed services can have the \textit{same} traffic, average latency, and errors, but different topological invariants, so they need different corrective actions. Topological invariants capture architectural constraints that remain hidden to standard performance metrics yet fundamentally shape system behavior. 
In particular, the harmonic component of the information flows between the functions is not generated by the load, but they are highlighted by the load which, therefore, allows corrective measures to be identified.

The paper is structured as follows. Section \ref{sec:related_work} discusses existing approaches and clarifies how our setting, based on a directed service graph whose nodes are FaaS functions, connects to TSP literature. 
Section \ref{sec:math_backgound} introduces the basic algebraic and topological concepts for our contribution. Section \ref{sec:problem_definition} defines the system assumptions and the addressed problem. Section \ref{sec:proposed_method} describes our proposal, with formal derivations. Section \ref{sec:experimental_results} shows the application of our proposal to two concrete case studies, quantifies metrics, and discusses the achievements. Finally, Section \ref{sec:conclusion} 
reports our concluding remarks and and outlines future work.

\section{Related Work} \label{sec:related_work}

\subsection{Combinatorial Hodge theory on graphs and edge flows}

The Eckmann's contribution on harmonic analysis on simplicial complexes~\cite{Eckmann1944} paved the way for discrete Hodge theory, establishing 
discrete analogues of differential forms, cohomology, and harmonic representatives. These ideas established the foundations for 
discrete exterior calculus (DEC), which provides the way to discretize differential operators (exterior derivative, Hodge star, and Laplacian) while preserving geometric structure~\cite{Desbrun2005DEC}. While DEC is often presented for meshes and manifolds, its operator viewpoint (incidence matrices as discrete derivatives) is aligned with graph-based Hodge decompositions used in network analysis.

The Hodge rank 
by Jiang \emph{et al.}~\cite{Jiang2011HodgeRank} 
models pairwise comparisons as an edge flow and decomposes it into gradient (globally consistent), curl (locally cyclic), and harmonic (global cycle) components; the relative magnitudes quantify different sources of inconsistency. Although the application domain is statistical ranking, the decomposition is generic for any edge-valued signal, and motivates our use of Hodge components to separate ``explainable'' function-to-function interactions from cyclic anomalies in service graphs.

Several works have broadened the applications of Hodge Laplacians and higher-order representations of networks. Horak and Jost~\cite{HorakJost2013} investigated spectra of combinatorial Laplace operators on simplicial complexes and established a commection between topological structure and eigenvalues, giving theoretical background for using Hodge Laplacian spectra as structural signatures. The Schaub \emph{et al.}~\cite{Schaub2020} work on higher-order network analysis developed diffusion and random-walk dynamics driven by Hodge Laplacians, showing 
how higher-order topology (beyond dyadic edges) affects mixing, bottlenecks, and flow patterns. These results are relevant for serverless services because real deployments often induce higher-order interaction patterns (e.g., fan-out/fan-in, join nodes, and multi-function coordination) that can be elevated from graphs to cellular complexes.

A recent paper from Schaub \emph{et al.} synthesize these concepts into practical guidance for data analysis, and provides a tutorial perspective on Hodge Laplacians for higher-order networks~\cite{Schaub2021Tutorial}, describing how to build boundary operators, interpret harmonic subspaces, and use Hodge-based filtering for edge and triangle signals. Our methodology adopts these constructions but tailors the signal definition to serverless observability, where edge weights can represent call counts, latency contributions, retry rates, or error propagation probabilities.
Barbarossa and Sardellitti \cite{Barbarossa2020TSP} extends signal processing from graphs to higher-order topological structures, allowing multi-node relationships to be analyzed using Hodge Laplacian operators. The extension 
to cell complexes is shown in \cite{Sardellitti}, where the authors propose a methods to infer a cell complex from data and alternative strategies to filter flow components.

\subsection{Serverless/FaaS computing}
Several position papers and surveys well define the serverless model and highlight key challenges. Baldini \emph{et al.}~\cite{Baldini2017Serverless} argue that serverless/FaaS is a shift toward event-driven, stateless functions with fine-grained billing, and outline open problems such as state management, debugging, performance predictability, and vendor lock-in. Hellerstein \emph{et al.}~\cite{Hellerstein2019Serverless} observe that serverless is a  paradigm shift for cloud programming, while also noting friction points (e.g., limited local control, performance variance, and new bottlenecks) that motivate better analysis and diagnostic tooling.

A significant body of empirical work focuses on performance issues and platform behavior, especially around cold starts, resource isolation, and multi-tenancy. Wang \emph{et al.}~\cite{Wang2018Peeking} perform a measurement study of serverless platforms and expose hidden performance issues, including cold start overheads and variability across providers. Such results directly motivate our graph-centric approach. In fact, if a service is represented as a function-invocation graph, performance anomalies can emerge as changes in edge flows (e.g., abnormal latency contributions along particular invocation edges), and Hodge decomposition provides a structured way to separate path-consistent slowdowns from cycle-driven effects such as retries and cascading failures.

For what concerns the architectural aspects, 
Akkus \emph{et al.} propose SAND~\cite{Akkus2018SAND}, a high-performance serverless runtime that reduces overheads via application-level sandboxing and more efficient communication among functions. They show that runtime architecture and inter-function communication can dominate end-to-end latency, suggesting the importance of modeling 
interactions as a network rather than treating each function in isolation.  
We argue that Hodge-based methods operate naturally on the interaction network (the function-invocation graph) and can incorporate architectural issues as perturbations to edge weights or higher-order structures.

In addition, there are emerging studies on real-world serverless workloads and the structure of serverless applications. Shahrad \emph{et al.}~\cite{Shahrad2020ServerlessInTheWild} analyze production traces to characterize invocation patterns, burstiness, and resource usage in serverless settings. Their findings suggest that function composition and workflow structure are central, and that graph/workflow representations are necessary for understanding system-level behavior. This supports our premise that a service graph should be treated as a first-class object for analysis.

For what concerns performance of serverless platforms, 
Bardsley et al. \cite{Bardsley2018} identify key evaluation issues, demonstrating how interconnected components induce latency in both basic architectures and sophisticated e-commerce deployments. 
Along the same line of research, Ngujen et al. \cite{hai2019} investigate performance guarantees for real-time serverless function deployment. Their work utilizes a mixed analytical-empirical model tailored for bursty, real-time applications across cloud and edge networks.

Copik et al. \cite{copik2021}  propose a comprehensive benchmarking framework, outlining a methodology to specify workloads, implementation, and infrastructure, which yields an abstract model of a FaaS execution environment. Complementing this, Bortolini and Obelheiro \cite {bortolini2020} empirically investigate performance and cost variations across FaaS providers, revealing a significant dependence on factors such as memory allocation, implementation language, and the provider itself.

With particular emphasis on management aspects, Wang et al. \cite{wang2021} introduce LaSS, a platform for latency-sensitive serverless services on edge nodes. It employs queuing-based methods for resource allocation and autoscaling to adapt to workload dynamics, incorporating a fair-share policy during overload. Similarly, Mittal et al. \cite{mittal2021} propose Mu, a Kubernetes-based platform that integrates core management components—including autoscaling, scheduling, and load balancing—with optimized management functions. Worker nodes propagate configuration and performance data to a central load balancer, enabling it to predictively route requests and proactively handle workload fluctuations.
Jia and Witchel \cite{jia2021} propose Nightcore, a serverless function runtime architecture including container-based isolation for functions and an efficient design that achieves microsecond-level latency for critical runtime operations.

Mahmoudi and Khazaei \cite{Mahmoudi2022} \cite{Mahmoudi2023} develop a performance model based on a Semi-Markov process, specifically targeting the analysis of cold starts. The model is experimentally validated, demonstrating accurate steady-state performance estimation. In the domain of node-level resource management, Panda and Sarangi \cite{Panda2024} design an intra-node manager that optimizes for latency and fairness by monitoring L1 cache, L2 cache, and CPU wait time, and propose a reinforcement learning strategy for dynamic priority and core distribution. On the system level, Ascigil et al. \cite{ascigil} present a FaaS model tailored for edge-cloud environments, comparing centralized and decentralized resource provisioning algorithms through event-based simulation. Complementing this, Pinto et al. \cite{pinto} formulate the function allocation problem between edge and cloud as a multi-armed bandit optimization.

\subsection{Bridging Hodge decomposition with 
FaaS observability}

To the best of our knowledge, there are no past research activities that explicitly refer to the combinatorial Hodge decomposition applied to serverless service graphs. However, a close conceptual match is HodgeRank-style decompositions~\cite{Jiang2011HodgeRank}, designed to identify inconsistency and cyclicity in edge data, as serverless traces naturally produce edge-valued measurements (calls, time, failures) on a function-call graph. Moreover, higher-order extensions via simplicial complexes~\cite{Schaub2020,Schaub2021Tutorial} can encode multi-function coordination units (e.g., a topological structure where multiple functions synchronize), 
difficult to represent with pairwise edges.

Our paper therefore contributes by (i) defining a principled mapping from serverless telemetry to edge/face signals; (ii) using Hodge decomposition to separate global trends from local and global cyclic behavior; and (iii) interpreting harmonic components as potential ``structural'' anomalies (e.g., persistent cycles corresponding to retries or compensating transactions) that are not captured by node-level metrics. This positions Hodge decomposition as a lightweight, explainable layer for serverless observability, complementary to black-box anomaly detection and to workflow-level optimization.

\section{Algebraic and Mathematical Background}  \label{sec:math_backgound}
The mathematical basis of the contribution presented in this paper consists in the representation of algebraic structures on oriented graphs. These algebraic structures can be either simplicial complexes or cellular complexes. In this paper we use cellular complexes because they are particularly well suited to modeling service deployment in serverless platforms, where not every combination of functions yields a valid service or saga. In contrast, a simplicial complex is closed under taking faces, which would require admissible combinations to be downward closed.

Let $V = \{v_0, \dots, v_{N-1}\}$ be a finite set of points, representing the
functions of a serverless service in what follows. 
A cellular complex $\mathcal{K}$ is a finite collection of cells of dimension
$k=0,1,2$, together with incidence relations specifying how higher-dimensional
cells are attached to lower-dimensional ones, satisfying the following
properties. Each $0$-cell corresponds to a vertex in $V$. Each $1$-cell is a distinct element $e_{ij} \in E$ associated with an ordered pair representing a logical relation between two $0$-cells $(v_i,v_j)$. Each $2$-cell, also referred to as $\xi \in F$, is attached along a closed chain of $1$-cells with a coherent orientation (e.g. counterclockwise). 
A \textit{service graph}, or \textit{function-invocation graph}, is defined as a graph $G=(V,E)$ including a cellular complex. 

On this topological structure, the spaces of the co-chains are defined as follows.
A $k$-cochain $C^k$ is a real-valued function on $k$-cells. In particular:

\begin{itemize}
\item $C^0=\{\alpha:V\xrightarrow{}\mathbb{R}\}$ - 0-cochain (function over nodes)  
\item $C^1=\{\omega:E\xrightarrow{}\mathbb{R}\}$ - 1-cochain (function over edges, with  $\omega(-e)=-\omega(e)$)
\item $C^2=\{\sigma:F\xrightarrow{}\mathbb{R}\}$ - 2-cochain (function over faces)  
\end{itemize}
Therefore:
    $\alpha\in C^0$  assigns a value to each vertex, conceptually similar to a potential, 
	$\omega\in C^1$  assigns a value, regarded as a flow, to each oriented edge. This function is arbitrary. For example it could be a function of the $\alpha\ $ value, or the mapping of values of experimental measures.
	$\sigma\in C^2$, assigns a value to each oriented face, associated with the information circulating between the functions that make up a saga or a whole service.

\textit{Differential operators} are linear maps that have the characteristic of increasing the degree of the chain to which the function belongs. In particular, the operator
\begin{equation}
d_0 : C^0 \to C^1
\end{equation}
\noindent has the role of calculating the discrete gradient of the graph. In fact, applying the operator to a function $\alpha$ defined on the vertices of an oriented edge $e=u \to v$, it returns 
\begin{equation}
(d_0 \alpha)(e) = \alpha(v) - \alpha(u)
\end{equation}
Therefore, the corresponding operation is a measure of the potential difference along the edge.
The differential operator 
\begin{equation}
d_1 : C^1 \to C^2
\end{equation}
\noindent can be associated with the operation of a discrete curl on the graph. In fact, considering an oriented face $\phi$ (e.g. counterclockwise) with edges $e_1,…,e_m$ coherently oriented, it returns:
\begin{equation}
(d_1 \omega)(\phi) = \sum_{i=1}^{m} \omega(e_i)
\end{equation}
Thus, the operation consists of a measurement of the net flow circulation around the face, i.e. the net load circulation between the functions associated with the face.

On each space we define a $C^k$ \emph{scalar product} in the usual form:
\begin{equation}
\langle \alpha, \beta \rangle_{C^k}
=
\sum_{c \in \{k\text{-cells}\}} \alpha(c)\,\beta(c).
\end{equation}

This definition was introduced to identify an \emph{adjoint operator} of $d_k$, also known as \textit{discrete codifferential}, as the linear operator

\begin{equation}
d_k^* : C^{k+1} \to C^k
\end{equation}
such that
\begin{equation}
\langle d_k \alpha, \beta \rangle_{C^{k+1}}
=
\langle \alpha, d_k^* \beta \rangle_{C^k},
\qquad
\forall \alpha \in C^k,\ \beta \in C^{k+1}.
\end{equation}
In particular,
\[
d_0^* : C^1 \to C^0
\]
is associated with the operator of the discrete divergence on the nodes of the
graph. In particular, given a vertex $v$, it is immediate to calculate
\begin{equation}
(d_0^* \omega)(v)
=
\sum_{\substack{\text{edges } e \\ \text{entering } v}} \omega(e)
-\sum_{\substack{\text{edges } e \\ \text{exiting } v}} \omega(e).
\end{equation}

\noindent
or with the opposite sign depending on the circulation convention. 

Using the above definitions, the Hodge Laplacian of a graph $L_k$ is defined as
\begin{equation}
L_k
=
d_{k-1} d_{k-1}^*
+
d_k^* d_k
:
C^k \to C^k .
\end{equation}
In particular, for $k = 1$:
\begin{equation}
L_1 = d_0 d_0^* + d_1^* d_1.
\end{equation}

A \emph{harmonic $k$-form} is a function $h \in C^k$ such that
\[
L_k h = 0.
\]
In other words, $h \in \mathcal{H}^k = \ker(L_k)$, i.e. the space belonging to the kernel of the Laplacian operator.

The Hodge theorem \cite{Jost2008} shows that

\begin{equation}
L_k h = 0
\iff
d_k h = 0
\ \text{and} \
d_{k-1}^* h = 0.
\end{equation}

Thus, a harmonic component is both a closed function
($d h = 0$) and a co-closed function ($d^* h = 0$).

The usefulness of using the Laplacian for graph analysis of functions derives
from its isomorphism with the $k$- group, defined as
\begin{equation}
H^k
=
\frac{\ker \bigl(d_k : C^k \to C^{k+1}\bigr)}
     {\operatorname{im} \bigl(d_{k-1} : C^{k-1} \to C^k\bigr)}.
\end{equation}
\noindent where $\operatorname{im} \bigl(d_{k-1}\bigr)$ denotes the image of the $d_{k-1}$ operator, and $H^k$ is a quotient vector space.
This consists closed $k$-forms modulo exact ones.
In particular, for $k=1$ the first cohomology group includes closed 1-forms,
i.e. independent cycles identifiable on the graph, which differ at most
by a gradient (i.e.\ a closed cycle that is the edge of a defined face).
Therefore, the $1$-th group of cohomology includes paths in the graph that
represent ``holes'', i.e.\ those cycles that cause problems and inefficiencies.

The fundamental result, which is the basis of our work, is the
\emph{discrete Hodge Theorem} \cite{Dodziuk1976} , which shows that
\begin{equation}
\mathcal{H}^k \cong H^k.
\end{equation}

The symbol $\cong$ in this context means canonical isomorphism between vector
spaces or groups.

This isomorphism between the space of harmonic co-chains, which belongs to the kernel of
$L_k$ and the $k$-th cohomology group is fundamental because every class contains \emph{exactly one} harmonic form.
Therefore, to study the holes in the logic of applications deployed in the
cloud we can use the spectral properties of the Laplacian.

Given a cohomology class \([\omega] \in H^k\), with $d_k \omega = 0$, there is only one \(h \in [\omega]\) such that:
\begin{equation}
d_k h = 0 \quad \text{e} \quad d_{k-1}^* h = 0
\end{equation}
Equivalently, $h$ is the unique harmonic element in the class satisfying \(L_k h = 0\). The discrete Hodge theorem yields the orthogonal decomposition:
\begin{equation}
C^k = \operatorname{im} \left( d_{k-1} \right) \oplus \mathcal{H}^k \oplus \operatorname{im} \left( d_k^* \right)
\end{equation}
\noindent where $ \oplus$ denotes the orthogonal direct sum of vector spaces.
Furthermore, it is worth to remember that the kernel size of Hodge Laplacian corresponds to the Betti Numbers, related to fundamental invariants \cite{Friedman1998}, which are those features that do not change when the geometry changes, depending only on the topology: 
\begin{itemize}
    \item $\beta_0 = \dim \ker L_0$, related to connected components. In particular, it indicates how many independent subsystems exist. In the serverless case, it indicates isolated functions, domains that do not communicate, real fault domains. Change only if you add or remove critical dependencies.
    \item $\beta_1 = \dim \ker L_1$, related to structural cycles. A nonzero value of $\beta_1$ indicates the number of non-contractable global loops. In the serverless case, reveals the presence of cyclic dependencies, which may lead to structural fragility or unavoidable retry behavior.
    \item $\beta_2 = \dim \ker L_2$, related to closed workflows, measures the number of nontrivial two-dimensional cohomology classes. 
In a serverless architecture, a nonzero $\beta_2$  reflects the presence of globally closed execution structures, multi-service transactions, or compensation-based workflows.
Topologically, these correspond to abstract “logical surfaces” in the dependency complex.
Beyond their count, the harmonic energy $\| h \|^2$, associated with a harmonic representative provides a quantitative measure of the global load carried by such structures. It is invariant under local rebalancing and increases only when the global architecture is structurally stressed.
\end{itemize}

It should also be noted that two flows \( f_1, f_2 \) can be regarded as equivalent if the harmonic component of $f_1-f_2$ is zero. 
In serverless settings, this means that two seemingly different, but structurally equivalent traffic patterns require the same architectural actions. For example, this can allow classifying incidents and group failure patterns.

Some useful information can also be deduced from the spectrum of the Laplacian. The eigenvalues of \( L_k \) are stable to local disturbances and change only with structural changes. If there are small gaps, the systems are fragile, while large gaps indicate robust systems.

In summary, the Hodge decomposition enables the separation of system-level metric flows into components arising from local imbalances (gradient terms), local cyclic interactions (curl terms), and global topological features that cannot be resolved by local adjustments.

\section{Problem Definition} \label{sec:problem_definition}

Consider an oriented graph graph with oriented vertices and edges. In the rest of the paper, each vertex will be associated with a deployed serverless function, and each edge connecting two nodes will be associated with the flow of information (broadly speaking) exchanged between the related functions. In what follows we refer to a saga as a sequence of transactions that maintains the consistency of data in different functions without distributed locks. Each function performs its own operation and initiates the next through events or messages; If a transaction fails, the saga behavior includes compensatory actions to undo the previous operations. 

Without compromising the generality of the model, we use a realistic running example used throughout the paper. It consists of a commercial application accessible online.  To be concrete, we consider the typical architecture of the Lambda environment \cite{AWSLambda}.
The classification and list of functions in Table \ref{tab:business-functions} are used to define sagas in the running example. Thus, a service (or saga) includes a certain number of nodes that constitutes another independent topological entity at a higher hierarchical level. This topological entity, which could be either a simplex or, more generally, a cell, corresponding to a $face$. 

To understand the depth of the proposed model, it is necessary to have a clear vision of what it means to organize an application in FaaS mode and what problems it could generate in a distributed deployment environment. 

We assume that the service functions are deployed in an infrastructure, which can be an edge cluster or a cloud datacenter, where communication latency between nodes (and thus functions) is negligible with respect to other service-related delays, such as service time, cold start delay, etc. 
A typical commercial application is broadly structured as follows:

\begin{itemize}
    \item Handler Function: It is the main function that contains the business logic. This is automatically invoked by the runtime of the deployment environment. Takes the event and context as parameters.
    \item Initialization functions, required for code started in cold mode: These are functions executed by the handler, only once for each new instance (cold start). These functions are used to initialize database connections, load configuration files, instantiate new complex objects.
    \item Core Business Functions, that characterize the implemented architecture with FaaS.
\end{itemize}

\begin{table}[t]
\centering
\caption{Core Business Functions for a Commercial Sample Application}
\label{tab:business-functions}
\footnotesize
\begin{tabularx}{\columnwidth}{p{1.8cm}lp{7cm}}
\toprule
\textbf{Layer/Category} & \textbf{Functions} \\
\midrule
\textbf{API Layer} & 
API Gateway + Lambda \\
\midrule
\multirow{3}{1.8cm}{\textbf{Core Infrastructure Functions}} & 
Authentication/Authorization (1--2 functions) \\
& Routing/Orchestration (1 function) \\\\
\midrule
\multirow{13}{1.8cm}{\textbf{Core Business Functions}} & 
\textbf{Product Catalog:} \\
& \quad getProducts -- Product list \\
& \quad getProductDetail -- Product detail \\
& \quad searchProducts -- Search \\
& \quad updateInventory -- Inventory Management \\
& \textbf{Shopping Cart:} \\
& \quad addToCart -- Add to cart \\
& \quad getCart -- View cart \\
& \quad updateCart -- Modify cart \\
& \quad clearCart -- Clear cart \\
& \textbf{Checkout \& Payments:} \\
& \quad initiateCheckout -- Start checkout \\
& \quad processPayment -- Process Payment \\
& \quad validatePayment -- Validate payment \\
& \quad handlePaymentWebhook -- Webhook payments \\
& \textbf{Orders:} \\
& \quad createOrder -- Create order \\
& \quad getOrderHistory -- Order history \\
& \quad getOrderStatus -- Order Status \\
& \quad cancelOrder -- Cancel order \\
\midrule
\multirow{8}{1.8cm}{\textbf{Background Functions}} & 
\textbf{Event Processing (SQS/SNS/EventBridge):} \\
& \quad processOrderFulfillment -- Order fulfillment \\
& \quad sendOrderConfirmation -- Email confirmation \\
& \quad updateRecommendations -- Update recommendations \\
& \quad syncInventory -- Sync Inventory \\
& \textbf{Cron Jobs (EventBridge Scheduler):} \\
& \quad generateDailyReports -- Daily reports \\
& \quad cleanupExpiredCarts -- Cart cleaning \\
& \quad backupDatabase -- Data backup \\
\midrule
\multirow{3}{1.8cm}{\textbf{Support Functions}} & 
\textbf{Analytics:} \\
& \quad logUserActivity -- Activity tracking \\
& \quad generateAnalytics -- Analytics \\
& \quad abTestHandler -- A/B testing management \\
\midrule
\multirow{3}{1.8cm}{\textbf{System Functions}} & 
\textbf{Administration:} \\
& \quad adminProductCRUD -- CRUD Products (admin) \\
& \quad viewSystemMetrics -- System metrics \\
& \quad processBulkUpload -- Bulk upload \\
\bottomrule
\end{tabularx}
\end{table}

Considering the level of deployment redundancy that characterizes the main deployment technologies, such as OpenFaaS over Kubernetes \cite{openfaasHome}, we arrive at the simultaneous deployment of a hundred functions. Considering that different instances of the same application can be implemented and that recycling them for different applications is considered a best practice, the intricacy of the logical relationships that are generated by making it interact is difficult to control.

Given a classic microservice architecture modeled as a cellular complex $\mathcal{K}$, a sequence of observed runtime flows $\{f(t)\}_{t=1}^T \subset C^1(\mathcal{K})$, the goal of this work is to:

\begin{enumerate}
    \item decompose observed flows using the Hodge decomposition;
    \item quantify the relative energy of gradient, curl, and harmonic components;
    \item identify structural performance bottlenecks associated with non-trivial harmonic flows;
    \item distinguish between locally correctable inefficiencies and globally structural issues.
\end{enumerate}

This formulation enables a principled, topology-aware analysis of microservice behavior that goes beyond traditional graph-based observability methods.

In serverless systems, dependencies are not always just between pairs of services, they often involve call loops, compound workflows, multiple requests across multiple functions. The proposed modeling consists in defining a cellular complex, of saga of functions, that characterize the service.  To accurately model multi-service execution pipelines without enforcing artificial simplicial closure, we adopt a cellular complex representation. This choice preserves the algebraic structure required for Hodge decomposition while allowing semantically meaningful higher-order interactions.

For a clear description of the model, we clarify what is \textit{not} a topological invariant, e.g. metrics such as requests per second (RPS), mean latency, CPU usage, number of pod. In fact, all of these can change without changing the topology. Topological invariants provide architecture-level guarantees that remain unchanged under local traffic fluctuations, enabling a principled distinction between operational and structural causes of system inefficiency. 

\section{Proposed Method} \label{sec:proposed_method}

For the definition of a topological structure of a serverless system, let $\mathcal{K} = (K_0, K_1, K_2) $ be a cell-oriented complex modeling system, representing:

\begin{itemize}[leftmargin=*]
    \item $K_0$: 0-cells, functions (nodes)
    \item $K_1$: 1-cells, directed calls between functions (RPC/HTTP)
    \item $K_2$: 2-cells workflow of 3 or more functions
\end{itemize}

In particular, $K_2$ represents interactions that cannot be reduced to pairs, typical of function pipelines.

Observable flows must be defined. For example, number of calls per function, the flow of requests, number of errors/retries per function, or other. These flows can be used to define 1-cochains (flows on edges) and 2-cochains (flows on cells).

The support of the cochains defined in Section \ref{sec:math_backgound} are:

\begin{itemize}[leftmargin=*]
    \item $C^0(\mathcal{K}) : \mathbb{R}^{|K_0|}$ - Values on functions
    \item $C^1(\mathcal{K}) : \mathbb{R}^{|K_1|}$ - Flows over edges
    \item $C^2(\mathcal{K}) : \mathbb{R}^{|K_2|}$ - Cell workflows
\end{itemize}

Every observable quantity (latency, traffic, errors) can be a vector in one of these spaces.
In the discrete domain deriving from the proposed cell complex, we construct the discrete operators, corresponding to the generic ones presented above  making use of the incidence relations of the edges on the nodes. Similarly to \cite{Schaub2021Tutorial} and \cite{Barbarossa2020TSP}, the edge operators take the form of oriented incidence matrices:

\begin{itemize}[leftmargin=*]
    \item $B_1 \in \mathbb{R}^{|K_0| \times |K_1|}$ (nodes $\to$ edges)
    \item $B_2 \in \mathbb{R}^{|K_1| \times |K_2|}$ (edges $\to$ faces or cells)
\end{itemize}

For the construction of incidence matrices, given the cellular complex $\mathcal{K}$ it suffices to enumerate functions (index set $K_0$), enumerate directed service interactions (index set $K_1$) and enumerate higher-order interaction motifs (index set $K_2$).

\begin{equation}
(B_1)_{v,e} =
\begin{cases} 
+1 & \text{if } v \text{ is the head of } e \\
-1 & \text{if } v \text{ is the tail of } e \\
0 & \text{otherwise}
\end{cases}
\end{equation}

\noindent $B_2$ is defined analogously using orientation consistency. Both matrices are sparse by construction. 
It immediately emerges that $B_1 B_2 = 0$.
This identity embodies the essence of the border functions, and is the topological basis of the cycles since ensures consistency between local interactions and higher-order structures. $B_1$ encodes the relationship between functions and calls, and $B_2$ encodes the relationship between calls and higher-order workflows. With reference to FaaS systems, $B_1 x$ it represents the divergence of any call flow $x$, which models the load imbalance by function. $B_1^T x$ represents the flow circulation, so loops and cyclic retries.

Hodge Laplacians in matrix form are defined as follows:
\begin{equation}
L_0 = B_1 B_1^T
\end{equation}
\begin{equation}
L_1 = B_1^T B_1 + B_2 B_2^T
\end{equation}
\begin{equation}
L_2 = B_2^T B_2
\end{equation}
\noindent We focus on $L_1$ because function call flows are modeled by 1-cochains. Let it therefore be $f \in C^1(\mathcal{K})$ an observed flow (e.g., calls, error rate). Since we are interested in the analysis of flows on the graph, the structural problem is:
\begin{equation}
L_1 x = f
\end{equation}
\noindent where $x$ is the potential generating (or correcting) $f$, the runtime observation. According to Hodge decomposition every flow $f\in C^1$ admits an orthogonal decomposition:
\begin{equation}
f = B_1^T \phi + B_2 \psi + h 
\end{equation}
\noindent with $\phi \in C^0$, $\psi \in C^2$, $h \in \text{ker } L_1$. $h$ is the part that no local balance can eliminate. This harmonic component satisfies the conditions $B_1 h = 0$ and $B_2^T h = 0$.

This means that no harmonic function is source or sink, no local loop explains the flow, and the problem is topological and global, at least at the level of sagas. Therefore, the harmonic component of the Hodge decomposition can capture the structural inefficiencies in serverless architectures that cannot be mitigated through local load balancing or retry control mechanisms.

We assume that the system topology is fixed over the observation window, flows are aggregated over time intervals sufficient to ensure stationarity, and function interactions are inferred from execution traces. Relaxing these assumptions and extending the framework to dynamic topologies is left for future work.

\subsection{Procedure for Hodge Decomposition of Serverless Flows}

In FaaS systems there is no flow conservation. An incoming request in a function can be processed and terminated (well), generate $N$ output requests (amplification), or generate $0$ outbound requests (absorption). However, we can use metrics on edges that have the meaning of ``flow'', even if not conservative. The detailed procedure is as follows.

\subsubsection{Construct $B_1, B_2$} Done by using the known service functions and APIs. 

\subsubsection{Solve $L_0 \phi = B_1 f$}
%
In a real FaaS system, the flow of requests has $f$ sources, i.e. entry points that receive external requests (net incoming flow $> 0$), it has sinks, i.e. functions that do not call other functions (net outgoing flow $> 0$), so it does not belong to $\operatorname{im}(B_1^T)$ (the space of conservative flows). In addition, measurement errors and stochastic behavior may occur. We want to find $\phi$ such that $f_{\text{grad}}$ is the orthogonal projection of $f$ on the space $\operatorname{im}(B_1^T)$ (conservative flows). The gradient component lies in $\operatorname{im}(B_1^T)$ and is obtained by solving:
\begin{equation}
\min_\phi J(\phi) = \| f - B_1^T \phi \|^2 = f^\top f - 2f^\top B_1^T \phi + \phi^\top B_1 B_1^\top \phi,
\end{equation}
Thus, deriving with respect to $\phi$, $\nabla_\phi J(\phi) = -2B_1 f + 2B_1 B_1^\top \phi = 0$ the normal equations $B_1 B_1^\top \phi = B_1 f$ are obtained, since $L_0 \phi = B_1 f$ and $L_0 = B_1 B_1^\top$. $L_0$ is singular, as $L_0$ always has the zero eigenvalue with eigenvector $\mathbf{1} = (1, 1, ..., 1)^\top$ ($B_1^\top \mathbf{1} = 0$ because each column of $B_1$ has zero sum: $+1$ and $-1$). Therefore, the system $L_0 \phi = B_1 f$ has a solution only if $B_1 f$ is orthogonal to $\mathbf{1}$ (compatibility condition), i.e. $\mathbf{1}^\top (B_1 f) = 0$. But since $\mathbf{1}^\top B_1 = 0$ by construction, the system is solvable (but with infinite solutions).
To make the solution unique, we used the Moore–Penrose pseudoinverse $L_0^+:$ 
    \begin{equation}
        \phi = L_0^+(B_1 f)
    \end{equation}

\subsubsection{Compute the gradient flow} $f_{\mathrm{grad}} = B_1^\top \phi$.

\subsubsection{Solve $L_2 \psi = B_2^\top f$}
The curl component lies in $\mathrm{im}(B_2)$ and is obtained by solving:
\begin{equation}
    \min_{\psi} \| f - B_2 \psi \|^2
\end{equation}
The normal equations give $B_2^\top B_2 \psi = B_2^\top f$, that is $L_2 \psi = B_2^\top f$, where $L_2 = B_2^\top B_2$.

\subsubsection{Compute  $f_{\mathrm{curl}} = B_2 \psi$} 
This system is typically much smaller than the gradient system, since $|K_2| \ll |K_1|$. The curl flow is $f_{\mathrm{curl}} = B_2 \psi$.

\subsubsection{Compute $h = f - f_{\mathrm{grad}} - f_{\mathrm{curl}}$} 
The harmonic component is the residual $h = f - f_{\mathrm{grad}} - f_{\mathrm{curl}}$. Equivalently, $h$ is the orthogonal projection of $f$ onto $\ker(L_1)$.  
In practice, the residual computation is numerically stable and avoids explicit nullspace computation.

The computational complexity is easily found as follows. Let $n = |K_0|$, $m = |K_1|$, $p = |K_2|$. Then:
\begin{itemize}
    \item constructing $B_1, B_2 : O(m + p)$.
    \item solving $L_0 \phi = B_1 f : O(m \log n)$ with modern Laplacian solvers.
    \item solving $L_2 \psi = B_2^\top f : O(p^3)$ (or faster if $B_2^\top$ is sparse).
\end{itemize}
\noindent The overall complexity is dominated by the gradient solve. Therefore, there are no particular scalability issues for the vast majority of serverless services. If the analysis is extended to higher levels, as at datacenter level, if the graph is becomes very large (thousands of functions), to solve $L_0 \phi = B_1 f$ it would be necessary to use iterative methods \cite{SpielmanTeng2014}.
Note that the energy of each component is invariant under orientation changes. If $K_2 = \emptyset$, the curl component vanishes and the model reduces to classical graph Laplacian analysis. The algorithm applies independently to each time window, enabling time-series analysis of component energies.

\subsection{Definition and analysis of flows}
In the simplest cases, based only on fewer always-on microservices, it would be sufficient to use oriented graphs without 2-cells. The incidence matrix $B_1$ from nodes to edges is easily found. The curl component refers to flows that are the sum of simple cycles: each cycle is a combination of 1-cycles, but in the flow field space, the projection on cycles is well defined by the dual of $B_1^\top$. In the FaaS case, when the number of deployed entities increases and the deployment process is dynamic, the proposed analysis allows you to identify behaviors that would otherwise be difficult to interpret:
\begin{itemize}
    \item \textit{Gradient component}: indicates flows generated by “potential” differences between nodes (e.g. a function calls another because the latter offers a resource, similar to a pressure gradient).
    \item \textit{Rotational component}: Could reveal call loops between functions (undesirable in ideal serverless architectures, but possible in complex orchestration scenarios). Useful for identifying circular dependencies.
    \item \textit{Harmonic component}: In a FaaS call graph, if the graph is acyclic and completely connected in a hierarchical way, the harmonic component may be zero; otherwise the orchestration may not be complete and harmonic components exist around the orchestration holes.
\end{itemize}

Main metrics are reported in Table  \ref{tab:topological-cycles-long}. The fields of the Table are detailed through the manuscript.
It should be considered that in general some metrics can be naturally mapped on nodes, i.e.\ associated with functions (e.g., CPU/Memory usage by the function, execution duration - internal processing time, cold start frequency, cost per execution) and some other metrics are mapped on edges, i.e. the calls (e.g., volume of requests between functions, network latency between functions, error rate of communication, dependency). Thus, we consider two types of variables:
\begin{itemize}
    \item $x \in \mathbb{R}^n$: Node metrics (CPU, memory, execution time)
    \item $y \in \mathbb{R}^m$: Edge metrics (network latency, communication errors).
\end{itemize}

In the case of edge-defined metrics, we can directly apply the approach described. In the case of node-defined metrics, we use Hodge decomposition to analyze co-chains defined on nodes, but instead of applying Hodge decomposition to raw metrics, we apply it to derived quantities. If we simply defined $f = B_1^\top x \Rightarrow f \in \mathrm{im}(B_1^\top)$ then for Hodge decomposition 
$f_{\mathrm{grad}} = f$ (all), $f_{\mathrm{curl}} = 0$, $f_{\mathrm{harm}} = 0$. So the algorithm would be trivial because it is by construction a purely gradient flow. Therefore, in this case it is convenient to define $f \in \mathbb{R}^m$ as a correlation or reciprocal influence between the functions based on the type of desired analysis, then apply Hodge decomposition to $f$. For example, to analyze how cold start frequency or errors affect the graph we can define:
\begin{equation}   
    f_{ij} = \frac{x_i - x_j}{\max(x_i, x_j)},
\end{equation}
\noindent
thus obtaining results such as:
\begin{itemize}
    \item $f_{\mathrm{grad}}$: CPU cores differences explainable by natural hierarchy;
    \item $f_{\mathrm{curl}}$: Cycles of imbalance (e.g. A $>$ B $>$ C $>$ A) indicating problematic control loops;
    \item $h$: Complex non-hierarchical patterns.
\end{itemize}

Similarly, to analyze how two functions vary together over time, useful for root cause analysis, we can use
\begin{equation} 
    f_{ij} = \mathrm{cov}(x_i(t), x_j(t)),
\end{equation}
\noindent thus obtaining results such as: 
\begin{itemize}
    \item $f_{\mathrm{grad}}$: Hierarchical correlations (master-slave);
    \item $f_{\mathrm{curl}}$: Circular correlations (A$\leftrightarrow$B$\leftrightarrow$C$\leftrightarrow$A) which indicates circular coupling;
    \item $h$: Complex correlations (Non-hierarchical clusters).
\end{itemize}
\noindent We can use weighted flows that combine call-relevant metrics with caller status, useful, for example, for cost analysis:
\begin{equation}
    f_{ij} = y_{ij} \cdot x_i,
\end{equation}
\noindent thus obtaining results such as:
\begin{itemize}
    \item $f_{\mathrm{grad}}$: “Natural” cost due to architecture
    \item $f_{\mathrm{curl}}$: Cyclical cost (circular calls that cost)
    \item $h$: Abnormal cost (e.g. expensive calls in unexpected routes)
\end{itemize}

After the analysis, different control actions can be applied. For example, reducing $f_{\mathrm{grad}}$ allows improved load balancing, reducing $f_{\mathrm{curl}}$ has an impact on retry and routing policies, reducing $h$ is achieved through architectural refactoring.

We stress that the topological invariants in serverless functions are quantities that do not change under load balancing, horizontal scaling, local variations in traffic. They change if you change the architecture, break or add dependencies, and redefine the workflow. \textit{Thus topological invariance is not a characteristic of performance, but of structure.}

The practical effects of the problems that may arise can affect practically all the main performance metrics that are shown in Table \ref{tab:topological-cycles-long}, and that can be used for the definition of co-chains for service analysis. 
In the table, $\beta_0 =1$ means causal/temporal cycle (1D). Only one component connected. It means that  there is only one zone and you stay in it, even when the action fails and you try again.
$\beta_1 =1$ means a non-reducible cycle (loop). It is a true cycle, which has been traveled to infinity. The $L_1$ harmonic captures its own energy trapped in the cycle, something that "spins" and does not break.
$\beta_2 =1$ means a closed surface / fork–join / saga (topological 2D). The graph does not only have cycles, but it has a closed surface, a 2D structure, i.e. there is a closed surface generated by compensations.
Harmonic compenent on $L_0$ means nodes/states that cannot be eliminated, it is a static problem, of "where energy get stuck". 
Harmonic compenent on $L_1$ concerns cycles. It is a dynamic problem, of "energy that turns". 
Harmonic compenent on $L_2$ concerns surfaces. It is a multidimensional problem, with potential multiple paths.

\begin{table*}[htbp]
\centering
\caption{Topological Characterization of Cycles in Distributed Architectures}
\label{tab:topological-cycles-long}
\renewcommand{\arraystretch}{1.2}
\scriptsize
\begin{tabularx}{\textwidth}{|
    >{\raggedright\arraybackslash}X |
    >{\raggedright\arraybackslash}X |
    >{\raggedright\arraybackslash}X |
    >{\raggedright\arraybackslash}X |
    >{\raggedright\arraybackslash}X |
    >{\raggedright\arraybackslash}X |
    >{\raggedright\arraybackslash}X |
    >{\raggedright\arraybackslash}X |
}
\hline
\textbf{Service / Saga Example} &
\textbf{Architecture -- logical flow (with explicit returns)} &
\textbf{Topological Model} &
\textbf{Cycle (Operational or Compensative)} &
\textbf{Betti invariants} &
\textbf{Nature of harmonic component} &
\textbf{$\beta$ reducible? } &
\textbf{Observable metrics} \\
\hline

E commerce &
(1) Orders $\rightarrow$ (2) Payments $\rightarrow$ (3) Shipping $\rightarrow$ (4) Inventory $\circlearrowleft$ rollback $\rightarrow$ (1) Orders &
Node = function \newline
Edge = RPC \newline
Cell = 2D workflow (e.g. checkout) &
Closed transactional loop &
$\beta_0 = 1$ \newline
Structural retry loop \newline
Strongly connected system &
Global harmonic component on $L_0$ &
No \newline
Just breaking the cycle (event driven / async) &
P99 end to end latency\newline
Retry rate \newline
Error amplification \\
\hline

Orchestration and FaaS (Saga) &
(1) API Gateway $\rightarrow$ (2) Validate $\rightarrow$ (3) Enrich $\rightarrow$ (4) Persist $\circlearrowleft$ compensate $\rightarrow$ (2) Validate &
Node = function \newline
Edge = events / RPC \newline
Cell = saga (2D), orchestration (3D) &
Compensation cycle &
$\beta_2 = 1$ \newline
Closed surface \newline
The saga generates a 2nd order constraint (not a classic cycle) distributed transaction &
Harmonic component on $L_2$ &
Partially \newline
By reducing compensations, async &
P99 / P999\newline \# compensations 
Retry step \\
\hline

ML Platform (with complex pipeline) &
(1) Ingest $\rightarrow$ (2) Preprocess $\rightarrow$ (3) Train $\rightarrow$ (4) Validate $\circlearrowleft$ feedback $\rightarrow$ (2) Preprocess &
Node = pipeline step \newline
Edge = data dependance &
Retraining feedback, \newline
The trained model generates new data that automatically restart part of the pipeline to retrain it. Real, non-compensative cycle. &
$\beta_1 = 1$ \newline
Global feedback &
High harmonic component on $L_1$ &
No \newline
It is not possible to elimicate the loop because the retraining is structural. \newline
Dampenable only (batching, accumulate updates) &
Loop period \newline
Queue depth \newline
Training latency \\
\hline

Cold start + distributed retries &
(1) Invoke A $\rightarrow$ (2) Cold start B $\rightarrow$ (3) Timeout $\circlearrowleft$ retry $\rightarrow$ (1) Invoke A &
Node = function \newline
Edge = invocation &
Causal cycle, a single connected component that causally recreates the same sequence &
$\beta_0 = 1$ \newline
Time cycle, not graph, so $\beta_1 = 0$. &
Local harmonic component on $L_0$ &
Yes (partial) \newline
Warm pool, backoff &
Cold start latency \newline
Retry rate \newline
P99 function \\
\hline

Fan out / fan in with asymmetric timeout &
(1) Orchestrator $\rightarrow$ (2) Fan-out (N calls) $\rightarrow$ (3) max wait $\rightarrow$ (4) Merge $\circlearrowleft$ timeout $\rightarrow$ (1) Orchestrator &
Node = function \newline
Edge = fork / join &
Partial closure. Asymmetric timeout creates a kind of ``surface'' in time / space &
$\beta_2 = 1$ \newline
Waiting for the maximum. \newline
Each branch is a 1-simplex, The ``Wait-for-All'' join creates an implicit 2-cell (waiting surface) &
Harmonic component on $L_2$, the join surface &
Partially \newline
Reduce fanout, async &
Max branch latency \newline
P99 merge - proceeds when all results are available or when a timeout expires \\
\hline

Eventual consistency backend &
(1) Write $\rightarrow$ (2) Read stale $\rightarrow$ (3) Detect inconsistency $\circlearrowleft$ compensate $\rightarrow$ (1) Write &
Node = state of data \newline
Edge = async &
Convergence loop &
$\beta_1 = 1$ \newline
loop not contractable, not eliminable without sacrificing any consistency &
Harmonic component on $L_1$. &
No \newline
Trade off with consistency &
Staleness window \newline
Update queue lag to be processed \\
\hline

Rate limiting provider &
(1) Request $\rightarrow$ (2) Throttle (429) $\rightarrow$ (3) Backoff $\circlearrowleft$ retry $\rightarrow$ (1) Request &
Node = endpoint \newline
Edge = External control &
Rate limiting (external) cycle. &
$\beta_1 = 1$ \newline
Load trapped by an ``external control loop'' &
Harmonic component on $L_1$ &
No \newline
Also using circuit breakers, limiting concurrency, using local token buckets, $\beta_1$ persists. &
Rate of HTTP 429 --- "Too Many Requests" response messages \newline
Retry storm \\
\hline

Access to incoherent caches &
(1) Read cache A $\rightarrow$ (2) Update source $\rightarrow$ (3) Invalidate async $\rightarrow$ (4) Read cache B (stale) $\circlearrowleft$ invalidate $\rightarrow$ (1) Read cache A &
Node = cache \newline
Edge = invalidation &
Invalidation loop. Messages (Kafka, Pub/Sub) that arrive with uneven delays, and nodes can receive updates in different order. &
$\beta_2 = 1$ \newline
Asynchronous invalidations create a surface of misaligned states \newline
The cycle is 2 dimensional &
Harmonic component on $L_2$ \newline
$L_2$ represents faces (2 cells), and Multi-node cache inconsistency is just a ``drift surface'' &
Partially, e.g. shortening TTL, using write through, ot server-side invalidation, but inconsistency persists due to the CAP theorem. &
Hit ratio \\
\hline

Partially idempotent workflow &
(1) Execute F1 (idempotent) $\rightarrow$ (2) Execute F2 (no idempotent side-effect) $\circlearrowleft$ retry $\rightarrow$ (1) Execute F1 &
Node = function \newline
Edge = side effect &
Causal cycle, F2 is not idempotent, produces duplication, sideeffect, side effects. Retry creates a causal loop that can generate duplication &
$\beta_0 = 1$ \newline
Duplications \newline
There is no topological cycle between nodes, but a causal cycle, introduced by the logic of retry. &
Harmonic component on $L_0$. 
A node that is re-executed creates side-effects. \newline
$L_0$ indicates that it is not a problem of cycles, but of nodes &
Yes \newline
Complete Idempotency \newline
Compensations &
Duplicate ops \\
\hline

Autoscaling based on threshold &
(1) Load spike $\rightarrow$ (2) Timeout $\rightarrow$ (3) Retry $\circlearrowleft$ feedback $\rightarrow$ (1) Load spike &
Node = group \newline
Edge = feedback &
Global feedback &
$\beta_1 = 1$ \newline
The cycle is real, a loop of instability: the system feeds itself, it is a negative control loop &
Harmonic component on $L_1$ \newline
$L_1$ indicates that the system continues to scale and descales, generating instability &
Partially \newline
Predictive scaling &
Scale oscillation \newline
P99 spikes \\
\hline

\end{tabularx}
\end{table*}

\section{Experimental Evaluation} \label{sec:experimental_results}
\begin{figure}[ht!]
    \centering
    \includegraphics[width=0.5\textwidth]{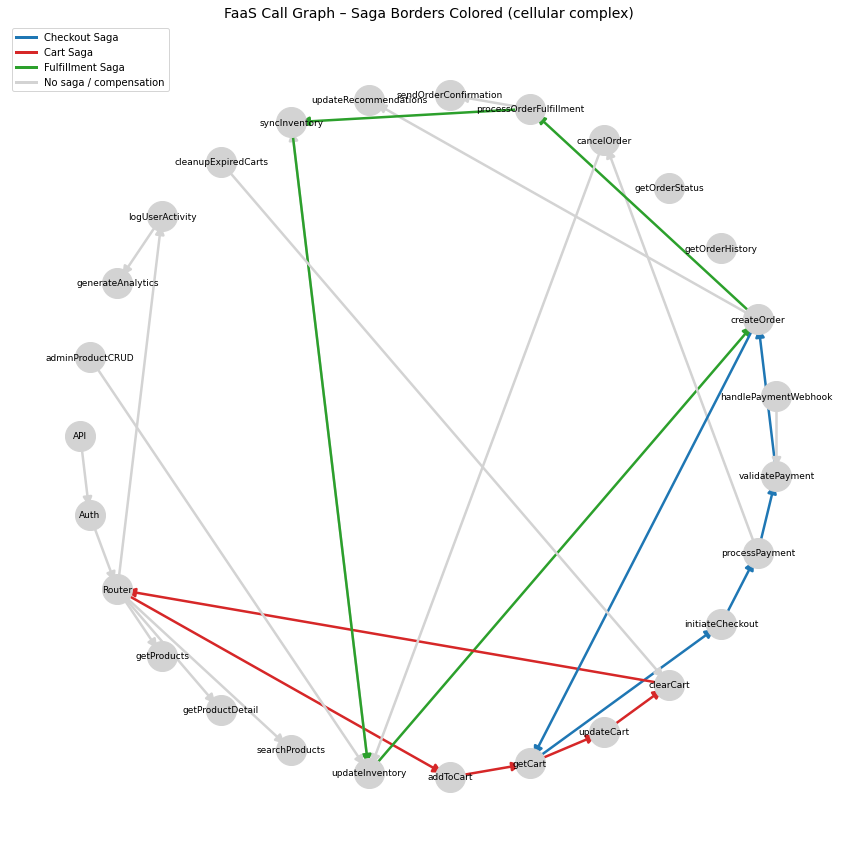} 
    \caption{Graph of the running example, including the defined sagas.}
    \label{grafo_servizio} 
\end{figure}
To obtain numerical results that show the potential of our proposal we make use of the running example. In addition to the functions shown, we introduce a compensation loop that is not handled like a saga.
In the graph of the service, shown in Figure \ref{grafo_servizio}, two functions are deliberately left isolated (\textit{getOrderStatus} and \textit{getOrderHistory}, on the top right), since they are not part of the core functions of the user interaction phase, and are called at a later time after service provision. Moreover, we can demonstrate the effectiveness in detecting isolated components, since these isolated functions emerge from the fact that they contribute to the introduction of additional null eigenvectors of $L_0=B_1B_1^T$, increase the kernel size of $L_0$ and do not participate in the flows identified in the Hodge decomposition. However, although they are invisible to decomposition, they emerge in the Betti numbers. The graph in Figure \ref{grafo_servizio} highlights the sagas defined over the service graph. Clearly, when such a service is deployed and orchestrated, a greater number of sagas is typically used. This example allows us to highlight the detection capabilities of the proposal.

\begin{figure}[ht!] 
    \centering
    \includegraphics[width=0.5\textwidth]{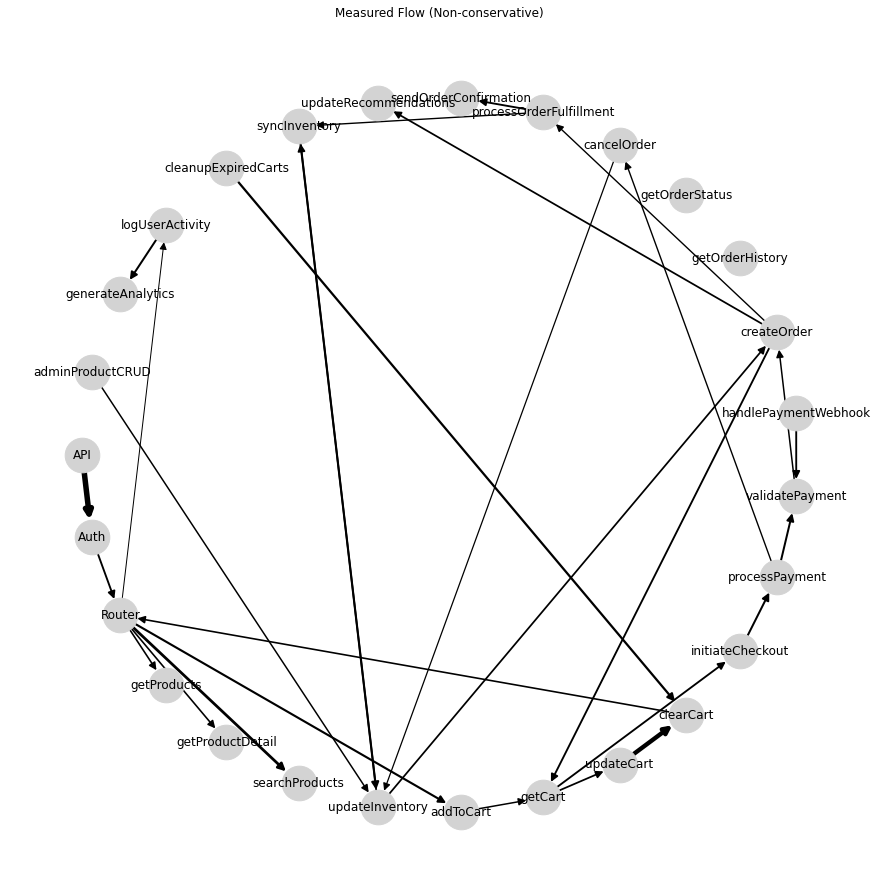} 
    \caption{Flow over edges of the invocation graph of the running example.}
    \label{sample_flow} 
\end{figure}

\subsection{Abnormal call flow}
In this first case study we analyze how traffic can stress the system structure. This analysis can be related to some of the problems reported in Table \ref{tab:topological-cycles-long}, such as asymmetric fan-out/fan-in generating some peaks of requests in some edges or autoscaling thresholds overcome by traffic peak. For this purpose, a co-chain \textit{f} has been associated with the graph. It is defined as a non-conservative flow of requests on the edges of the FaaS graph. Each component of \textit{f} represents the number of function invocations through edges measured on an observation time $T$, randomly generated for this analysis according to a Poisson distribution with an mean of $\lambda=10$ requests/$T$. For the purpose of our analysis, the load was increased in some edges. Specifically, the edge $API\rightarrow Auth$ received an increment of $\Delta\lambda=30$ requests/$T$, simulating massive traffic ingress from the API front-end. The edge $processPayment \rightarrow validatePayment$ received an increment of $\Delta\lambda=15$ requests/$T$, simulating flow amplification in some critical functions. In this way, \textit{f} becomes non-conservative: Some nodes (sources and sinks) have net flow in or out, just as it could happens in deployed FaaS systems.

\begin{figure}[h!] 
    \centering
    \includegraphics[width=0.5\textwidth]{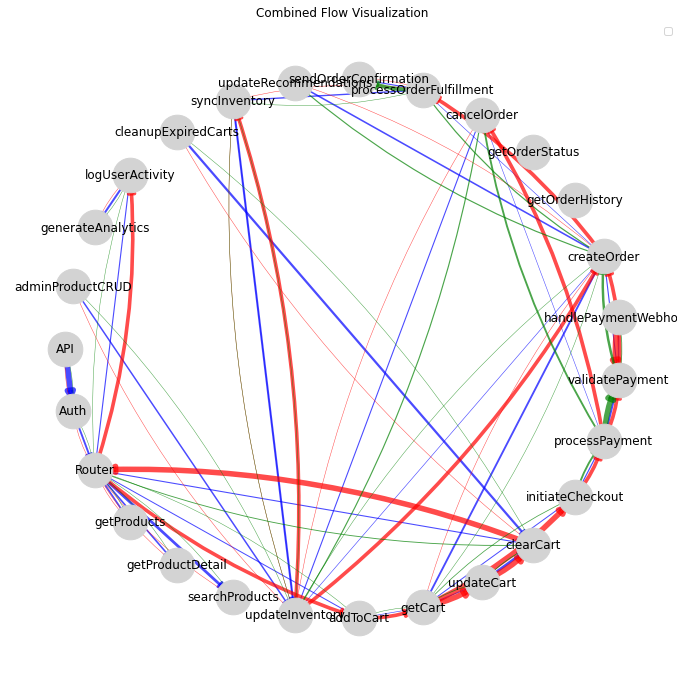} 
    \caption{Graph related to the function pressure case study. The gradient component in shown in blue, the curl component in red, and the harmonic component in green.The thickness of the edges for each color represents the weight of the component within the same decomposed component, not in absolute terms.}
    \label{base_combinata} 
\end{figure}

We could certainly insert many other edges to model all the problems illustrated in the Table \ref{tab:topological-cycles-long} in the running example, but, without loss of generalities, we limit ourselves to including an unmanaged compensation case study, capable of generating results that can be interpreted from an intuitive point of view. 
The sample measured flows over the graph is illustrated in Figure \ref{sample_flow}. Whereas, the Hedge decomposition of the co-chain is illustrated in Figures \ref{base_combinata}.

\begin{figure*}[h!] 
    \centering
    \includegraphics[width=0.75\linewidth]{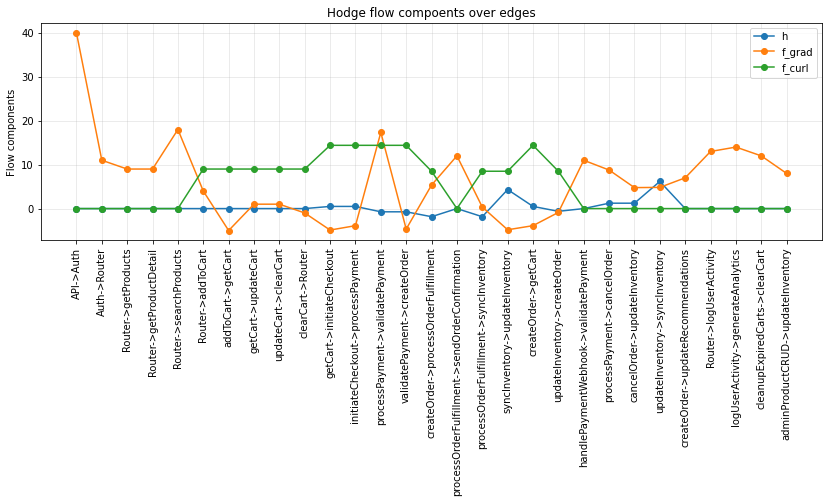} 
    \caption{Gradient, flow, and harmonic components over edges generated by the load modeled over the service graph. On the ordinate the unit of measurement is request/$T$.}
    \label{hodge_cold_lineare}
\end{figure*}

Figure \ref{hodge_cold_lineare} shows the Hodge flow components over edges. The figure reports the name of the edges, although in this figure the perception of the cyclical nature of the saga is lost, as delays of gradient and curl components are expected to be managed by sagas, whilst the large harmonic components clearly indicate the edges requiring specific interventions. 

The different pressure over edges clearly generates a significant gradient components. However, the defined sagas push part of the flow through curl components. The non-zero harmonic component quantitatively identifies persistent request circulation patterns that are divergence-free but not induced by any saga, revealing topological debt in the orchestration layer.
Betti's numbers for the invocation graph shown in Figure \ref{grafo_servizio} are: $\beta_0=3$, $\beta_1=3$, $\beta_2=0$.  
These results show that a non-trivial harmonic component in microservice call flows provides a mathematically grounded indicator of architectural inefficiencies that cannot be resolved through local control mechanisms.
The results obtained clearly show that the harmonic component his non-zero and is localized on a few edges, those that close compensation/retry cycles. 
$\beta_2 = 0$ is a consistent result, as definition of the case study does not include any information about possible 3D problems. In fact,  the sagas are 2-cell independent and no "sagas of sagas" have been introduced, and there are no redundant faces (no cavity size). 
$\beta_1 > 0$ is a diagnostic structural signal. In fact, it means that there are edge cycles that have zero divergence and are not edges of any saga. In the FaaS model, these loops can correspond to compensations, asynchronous retries, callbacks + webhooks, and non-explicit orchestration loops. 

It is clear from the results that $h$ is almost zero on most edges and is focused on the edges $cancelOrder \to updateInventory$, $processPayment \to cancelOrder$, webhooks/callbacks that are included into the flow. This means that the compensation cycle is active in real traffic, but it is not governed by a saga. In other words, the cycle is topologically "free". It is also worth highlighting the fact that the curl only captures loops that are edge of a face, whereas in the model the compensations are not sagas, so they are not free faces. Therefore, the result on the harmonic components means that the model is correctly detecting real unorchestrated compensation cycles.
The results depend directly on the values assigned to $f$, but the spatial structure of the decomposition (where \textit{gradient}, \textit{curl}, $h$ end) depends only on the graph and the sagas. Therefore, changing $f$ only changes the magnitudes, but it does not change the position of the harmonic cycles, nor the fact that an edge belongs to $f_{curl}$ or $h$. Therefore, this type of analysis provides valuable information on the structuring of the service, regardless of how the service is loaded at the operational level.

\subsection{Presence of Cold Start}
The analysis of cold start phenomena in the deployed service requires moving from a structural view  of the service architecture to a dynamic and performance reading.
Essentially, the presence of cold starts impacts the cost of traversing an edge or node in terms of latency. Therefore, cold start is a property of nodes, not edges, but the effect manifests itself on the incoming/outgoing edges of that node. 

The basic idea is to model the cold start using a weighted 1-cochain that measures the invocations affected by cold start, rather than the number of reciprocal calls between functions. We therefore introduce a 1-cochain flow $f^{cs}$ whose value increases with the number of cold functions and, consequently, as the additional average latency on the edges increases. More specifically, if the target function is in cold start then all incoming edges receive a high weight, while if it is warm the relative weight is 0. Of course, the effect only manifests itself if the function is called. For this reason, we combined the cochain that models the flow of requests with the cochain related to the cold start by means of the point product:

\begin{equation}
    f^{lat} = f^{req} \odot f^{cs}
\end{equation}

\noindent This co-chain can be decomposed using the Hodge decomposition, obtaining

\begin{equation}
    f^{lat} = f_{grad}^{lat} + f_{curl}^{lat} + h^{lat}
\end{equation}

\begin{figure}[h!] 
    \centering
    \includegraphics[width=\linewidth]{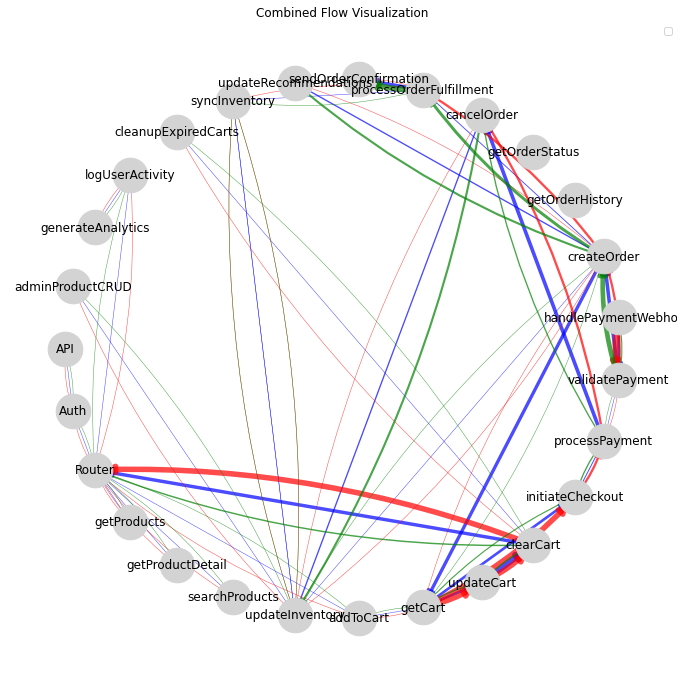} 
    \caption{Graph related to the cold start case study. The gradient component in shown in blue, the curl component in red, and the harmonic component in green. The thickness of the edges for each color represents the weight of the component within the same decomposed component, not in absolute terms.}
    \label{cold_combinata} 
\end{figure}

These results can be interpreted as follows. The gradient component \( f_{\text{grad}}^{\text{lat}} \) indicates latencies due to functional hotspots, i.e., functions that receive many calls from different points, such as API, Router, processPayment. Therefore, if a function is often cold and very central then it generates a dominant gradient. The curl component \( f_{\text{curl}}^{\text{lat}} \) indicates latencies that circulate within a saga and this implies that the cold start is "trapped" within the saga itself and does not depend on external inputs, but on the cycle itself. This is quantitative information of how much a saga is impacted by the cold start. Finally, the harmonic component \( h^{\text{lat}} \in \ker (B_1) \cap \ker (B_2^T) \) represents latencies that do not create accumulation of requests and are not explainable by any saga, such as compensation loops, asynchronous webhooks and retry paths. Hence, if \( h^{\text{lat}} \) is large there are systemic bottlenecks; if \( f_{\text{curl}}^{\text{cs}} \) is large there are sagas to "pre-warm"; if \( f_{\text{grad}}^{\text{lat}} \) is large there are functions to keep active at all times. 
In the running example the chain processPayment \(\rightarrow\) cancelOrder \(\rightarrow\) updateInventory \(\rightarrow\) syncInventory generates this effect. 
For example, we assume that  the functions \textit{processPayment}, \textit{validatePayment} and \textit{syncInventory} are cold, and model the additional latency with numerical values equal to 300, 200 and 400 ms, respectively. Figure \ref{cold_combinata} shows the decomposition of \( f^{\text{lat}} \).

Given their nature, to address problems arising from harmonic components, it is necessary to adopt "damping" solutions. For example, one could use strategic pre-warming, implement connection pooling to reduce the resonant effect of sequential calls, introduce adaptive timeouts that act as buffers in the flow, or use circuit breakers, which interrupt the harmonic propagation of errors.

\begin{figure*}[ht!] 
    \centering
    \includegraphics[width=0.75\linewidth]{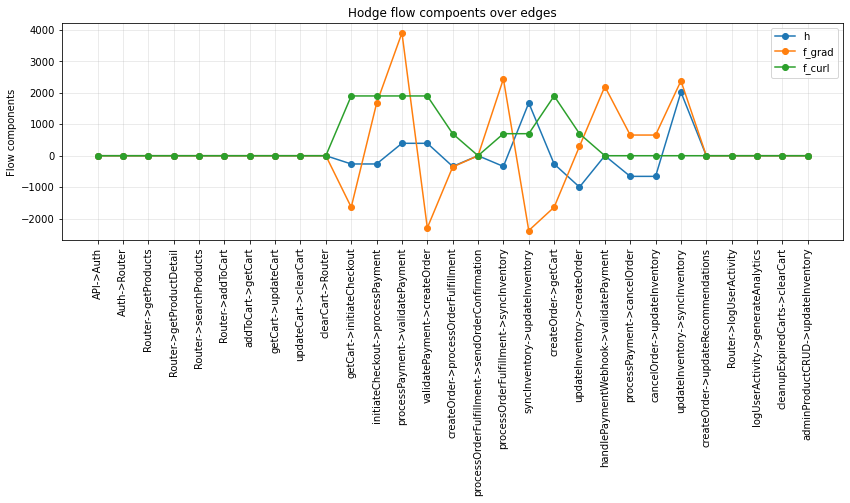} 
    \caption{Gradient, flow, and harmonic components over edges generated by cold start over three selected nodes. The metric shown in the ordinate axis highlight the impact of cold start over the flow components.}
    \label{hodge_cold_combinata} 
\end{figure*}

Figure \ref{hodge_cold_combinata} shows the Hodge flow components over edges. The figure shows the name of the edges in order to help the analysis of results. In particular, 
the figure shows which edges are affected by the fact that the hypothesized functions were found cold upon invocation.
Although in this representation the perception of the cyclical nature of the saga is lost, delays of gradient and curl components are expected to be managed by sagas. We clearly assume that sagas, along with their control actions,are properly defined. Instead, the large harmonic components clearly indicated the edges requiring specific interventions.

\section{Conclusion} \label{sec:conclusion}
Services deployed in serverless platforms typically generate flow information components, associated with interactions of functions,  that can be  of different nature. We identified such components by using the Hodge decomposition, that separates operational flows into locally correctable components and globally persistent harmonic modes.
In particular, harmonic flow appear since  in FaaS operation it is difficult to control control positioning and fine concurrency, and the system reacts by thresholds. 
These phenomena are not bugs, they are not configuration errors, but they are inevitable consequences deriving from the deployed serverless model and the topology makes them visible, and does not create them.
To correct any unexpected deviations in the metrics, in this paper we show a procedure capable of thoroughly analyzing the information flows exchanged by the deployed functions in order to identify the corrective action to be used, such as the adjustment of horizontal autoscaling, the introduction of appropriate corrections in the code of some functions, the correction of the hardware sizing, the adjustment of the control plane,  change cache policy, and so on. Our numerical analysis shows the capability of identifying such latent topological structures and distinguish them from transient overloads.
For future developments we will include other topological models for identifying harmonic components emerging through the interaction of different services competing for resources in the same datacenter.

\bibliographystyle{IEEEtran}
\bibliography{references}

\end{document}